\documentclass[12pt]{article}
\usepackage{amsmath,amssymb,graphicx}

\def\mydate{July 6, 2006}
\def\ignore#1{{}}

\newcounter{sxn}

\newcounter{axn}

\date{}

\newdimen\mybaselineskip
\renewcommand{\baselinestretch}{1.25}

\renewcommand{\thefootnote}{\arabic{footnote}}

\newcommand{\beeq}{\begin{equation}}
\newcommand{\eneq}{\end{equation}}
\newcommand{\beqn}{\begin{eqnarray}}
\newcommand{\eeqn}{\end{eqnarray}}

\def\dd{\partial}
\def\la{\raise.16ex\hbox{$\langle$}\lower.16ex\hbox{}  }
\def\ra{\, \raise.16ex\hbox{$\rangle$}\lower.16ex\hbox{} }
\def\go{\rightarrow}

\def\onehalf{ \hbox{${1\over 2}$} }

\def\Tr{{\rm Tr \,}}

\def\eff{{\rm eff}}

\def\calL{{\cal L}}
\def\D{{\cal D}}

\def\diag{{\rm diag ~}}

\def\psibar{ \psi \kern-.65em\raise.6em\hbox{$-$} }
\def\psibarl{ \psi \kern-.65em\raise.6em\hbox{$-$} \lower.6em\hbox{} }

\begin{document}
\thispagestyle{empty}

\baselineskip=12pt

{\small \noindent \mydate    \hfill OU-HET 563/2006}


\baselineskip=35pt plus 1pt minus 1pt

\vskip 3.cm

\begin{center}
{\Large \bf All-order Finiteness of the Higgs Boson Mass}\\
{\Large \bf in the Dynamical Gauge-Higgs Unification}\\

\vspace{3.0cm}
\baselineskip=20pt plus 1pt minus 1pt

{\def\thefootnote{\fnsymbol{footnote}}
\bf  Yutaka\ Hosotani\footnote[1]{hosotani@phys.sci.osaka-u.ac.jp}}\\

\vspace{.3cm}
{\small \it Department of Physics, Osaka University,
Toyonaka, Osaka 560-0043, Japan}\\
\end{center}

\vskip 2.5cm
\baselineskip=20pt plus 1pt minus 1pt

\begin{abstract}
In the dynamical gauge-Higgs unification, it is shown that
 the mass of the Higgs boson (4D scalar field) in $U(1)$ gauge
theory in $M^4 \times T^n$ ($n=1,2,3,\cdots$) is finite to all order 
in perturbation theory as a consequence of the large gauge invariance.
It is conjectured that  the Higgs boson mass is finite   
in  non-Abelian gauge theory  in $M^4 \times S^1$,   $M^4 \times (S^1/Z_2)$ 
and the Randall-Sundrum  warped  spacetime  to all order in the rearranged
 perturbation theory where the large gauge invariance is maintained.
\end{abstract}

\newpage


\newpage

In the standard model of electroweak interactions  the Higgs scalar field
plays a crucial role in inducing  the electroweak symmetry breaking, giving
finite masses to the weak bosons, quarks and leptons.    
Fundamental  scalar fields  are necessary  in the minimal 
standard supersymmetric model
(MSSM) and grand unified theories (GUT) as well.    Interactions associated with  
the Higgs fields, however,  are largely unconstrained.

The Higgs boson in the electroweak 
interactions is expected to be discovered at LHC (Large Hadron Collider)
in the near future.  
We are facing the time when the Higgs sector in elementary particles
is  disclosed where the key structure in the symmetry breaking hides.  
On the theoretical side the interaction of the Higgs field poses a challenging
problem concerning the stability of the Higgs boson mass against radiative
corrections.  It has been often argued that the Higgs boson mass suffers from 
quadratically divergent radiative corrections unless protected by symmetry.  
Supersymmetry provides desired protection, 
providing a leading candidate for physics beyond the standard model.
MSSM, in particular, 
predicts a light Higgs boson with a mass $m_H < 130$ GeV.  The experimental
lower bound from the direct search for the Higgs boson is 114 GeV.\cite{HiggsExp}

Recently an alternative scenario has attracted  attention from the viewpoint of
unification and stabilization of the Higgs field.   The Higgs field in four dimensions 
is unified with gauge fields within the framework of higher dimensional gauge
theory.   Low energy modes of 
extra-dimensional components of gauge potentials serve as 4D Higgs scalar fields.
Many years ago Fairlie and Manton proposed such unification scheme in six dimensions
compactified on $S^2$ by ad hoc symmetry ansatz.\cite{Fairlie1, Manton1}
Justification of the ansatz by quantum dynamics was also discussed.\cite{YH3}
A few years later it was  found that more natural scheme justified
by dynamics is provided when the extra-dimensional space is non-simply 
connected.\cite{YH1, YH2}
There appear Yang-Mills Aharonov-Bohm phases, $\theta_H$, associated with
the gauge field holonomy, or the phases of Wilson line integrals along noncontractible
loops.  Classical vacua are degenerate with respect to values of $\theta_H$.  
The degeneracy is lifted by quantum effects, thus the quantum vacuum being
dynamically determined by the location of the global minimum of the effctive
potential $V_\eff (\theta_H)$.  In non-Abelian gauge theory the gauge symmetry 
can be dynamically broken, depending on the value of $\theta_H$.   This 
 is called the Hosotani  mechanism. 
Fluctuations of  the Yang-Mills AB phases $\theta_H$ in four dimensions 
correspond to 4D Higgs  fields.  Higgs fields are unified with gauge field
and the gauge symmetry is dynamically broken.  The scheme is called the
dynamical gauge-Higgs unification.

It was  shown that the $\theta_H$-dependent part of the effective potential
$V_\eff (\theta_H)$ is finite at the one-loop level, irrespective of a regularization
method employed.\cite{YH1, YH2, Gersdorff}
This is highly nontrivial, given the fact that gauge theory in
higher dimensions is non-renormalizable.  As the curvature of the effective potential
at the global minimum is related to the mass of the 4D Higgs boson, $m_H$,  
it implies that a finite $m_H$ is generated radiatively with its value independent of
the cutoff scale in non-renormalizable theory.  
Although higher dimensional, non-renormalizable gauge theory is employed, 
predictions obtained there have 
sound meaning which does not depend of the details of dynamics at the cutoff scale.

The dynamical gauge-Higgs unification has been applied to both GUT and
electroweak interactions.\cite{Lim2}-\cite{Panico2}
   In particular, the dynamical gauge-Higgs unification of
electroweak interactions in the Randall-Sundrum warped spacetime\cite{RS1}  gives 
many interesting predictions in the Higgs field and gauge field 
phenomenology.\cite{Pomarol2}-\cite{HNSS}
The Higgs mass is predicted in the range between 140 GeV and 280 GeV, and
the Kaluza-Klein mass scale is predicted in the range 1.5 TeV and 3.5 TeV.\cite{HM}
The universality of the weak gauge interactions is slightly broken, and 
the Yukawa couplings are significantly suppressed.\cite{HNSS} These features can be 
measured at LHC and the future linear collider.

It becomes an important issue whether or not the mass of the 4D Higgs boson 
remains finite against higher order radiative 
corrections.\cite{YHscgt2, Morris, Irges, Maru1}   If it does, the 
dynamical gauge-Higgs unification scenario gives robust predictions concerning 
the Higgs and gauge field phenomenology, justifying the use of non-renormalizable gauge 
theory in constructing a unified theory.    It would give an alternative to 
supersymmetric theories to describe physics beyond the standard model.  
An important step in this direction has been taken
by the author a few years ago in outlining a proof for the all-order finiteness of 
the effective potential $V_\eff(\theta_H)$.\cite{YHscgt2} 
As we will see below,     
the argument in ref.\  \cite{YHscgt2} is valid in QED in arbitrary dimensions,
whereas the case of non-Abelian gauge theory requires further elaboration.   

Recently Maru and Yamashita performed detailed two-loop computations of the 4D Higgs
boson mass in QED on $M^4 \times S^1$.\cite{Maru1}   Their result supports 
the all-order result for the effective potential in ref.\ \cite{YHscgt2}.  
In this paper we give a proof for the  all-order finiteness of the effective potential 
$V_\eff(\theta_H)$ in QED in $M^4 \times T^n$,  where $T^n$ is an $n$-torus.
We argue that the finiteness of the Higgs boson mass remains valid
in non-Abelian gauge theory on $M^4 \times S^1$, $M^4 \times (S^1/Z_2)$,
and  the Randall-Sundrum (RS) warped spacetime as well. 
\ignore{Here $S^1/Z_2$ is  an orbifold obtained from $S^1$ by parity 
identification (of $ y$ and $-y \,$) in the extra dimension.  The RS spacetime
has topology of $S^1/Z_2$.}  
The most crucial ingredient in the proof is the large gauge
invariance associated with $\theta_H$.

\bigskip
\leftline{\bf 1.  Yang-Mills AB  phases $\theta_H$}

In gauge theory defined on a non-simply connected space, a configuration of
vanishing field strengths  $F_{MN}=0$ does not necessarily imply  trivial.  
Consider  an $SU(N)$  gauge theory on $M^4 \times S^1$ with
coordinates   $(x^\mu, y)$.  Boundary conditions are given by
\beqn
&&\hskip -1cm 
A_M(x,  y +2 \pi R) =
U A_M(x,  y ) \, U^\dagger  ~~, \cr
\noalign{\kern 5pt}
&&\hskip -1cm 
\psi(x, y + 2 \pi R) 
  =  e^{i\beta} \, T[U] \psi(x,  y) ~~,
\label{BC1}
\eeqn
where $U \in SU(N)$.   $T[U] \psi= U \psi$ or   
$U \psi U^\dagger$ for $\psi$ in the fundamental or adjoint representation,
respectively. The boundary condition (\ref{BC1}) guarantees that the physics is 
the same at $(x,  y)$ and $(x,  y + 2 \pi R)$.
\ignore{The theory is specified with a set of boundary conditions $\{ U, \beta \}$. }

Under a gauge transformation
\beeq
A_M'
=\Omega  A_M \Omega^{\dagger}
 -{i\over g} \Omega \dd_M \Omega^{\dagger}
 ~~, 
\label{gaugeT1}
\end{equation}
$A_M'$ obeys a new set of boundary conditions 
\beqn
&&\hskip -1cm 
A_M'(x,  y +2 \pi R) =
U' A_M'(x,  y ) \, U'^\dagger  ~~~, \cr
\noalign{\kern 5pt}
&&\hskip -1cm
U' = \Omega(x,  y + 2 \pi R)\, U \, \Omega(x,  y)^\dagger ~, 
\label{BC2}
\eeqn 
provided  $\dd_M U' = 0$.  Gauge transformations $\Omega (x,y)$ which
preserve the boundary conditions (so that $U' = U$) represent the residual gauge
invariance.  A set of   eigenvalues 
$\{ e^{i\theta_1}, \cdots, e^{i\theta_N} \}$ ($\sum_{j=1}^N \theta_j = 0$)
of $P \exp \big\{ ig \int_0^{2\pi R} dy \, A_y  \big\} \cdot U$ is invariant 
under residual gauge transformations.   These phases $\theta_j$'s 
are Yang-Mills AB phases (holonomy phases) associated with a 
non-contractible loop in non-simply connected space which are collectively 
denoted  as $\theta_H$.\cite{YH2}
On flat space,  constant configurations $A_y$ give nontrivial $\theta_H$.
Although $\theta_H$ gives vanishing field strengths at the classical level,
it affects physics at the quantum level.

Boundary conditions on an orbifold $M^4 \times (S^1/Z_2)$ are given
by \cite{HHHK}
\beqn
&&\hskip -1cm 
\begin{pmatrix}  A_\mu \\ A_y    \end{pmatrix} (x, z_j  -y ) 
= P_j \begin{pmatrix}  A_\mu \\  -A_y    \end{pmatrix}  (x,  z_j + y ) 
\, P_j^\dagger  ~~, \cr
\noalign{\kern 5pt}
&&\hskip -1cm 
\psi(x, z_j - y )   =  \pm  \, T[P_j ] \gamma^5 \psi(x, z_j + y) ~~.
\label{BC3}
\eeqn
Here $(z_0, z_1)=(0, \pi R)$, $P_j \in SU(N)$ and $P_j^2 =1$.  
It follows that 
$A_M(x + 2\pi R) = U A_M(x,y) U^\dagger$ where $U= P_1 P_0$. 
The residual gauge invariance is given by $\Omega(x,y)$ satisfying 
\beqn
&&\hskip -1cm
P_j =\Omega(x,  z_j -  y) \, P_j    \, \Omega(x, z_j + y)^\dagger ~, \cr
\noalign{\kern 5pt}
&&\hskip -1cm
U =\Omega(x,  y +2 \pi R ) \,  U  \, \Omega(x,   y)^\dagger ~.
\label{newBC1}
\eeqn 

As $A_y$ has an opposite parity to $A_\mu$, there may or may not
exist $\theta_H$, depending on $P_j$.  We are interested in cases
where $\theta_H$ exists so that its fluctuation mode is identified as
a 4D Higgs field.  For instance, in the $SU(3)$ model  with
$P_j= \diag (-1, -1, 1)$,  the constant part of $(A_y^{13}, A_y^{23})$ 
forms $\theta_H$.\cite{Lim1}   Its four-dimensional fluctuations correspond to
the $SU(2)_L$ doublet Higgs field.  In the $SO(5) \times U(1)_{B-L}$ model
with $P_j = \diag (-1,-1, -1, -1, 1)$, $A_y^{j5}$ $(j=1,\cdots,4)$ form
the $SU(2)_L$ doublet Higgs field.\cite{Agashe2}

The Randall-Sundrum (RS) warped spacetime is given by 
\beeq
ds^2 = e^{-\sigma(y)} \eta_{\mu\nu} dx^\mu dx^\nu + dy^2
\label{RSmetric}
\eneq
where $\eta_{\mu\nu} = \diag (-1,-1,-1,1)$, $\sigma(y)=\sigma(y+2\pi R)$
and $\sigma(y) = k |y|$ for $|y| \le \pi R$.  In the $k \go 0$ limit
the RS spacetime becomes $M^4 \times (S^1/Z_2)$.
Gauge fields in the RS spacetime obey the same boundary conditions 
as in (\ref{BC3}).  In the $SU(3)$ model with $P_j= \diag (-1, -1, 1)$, 
for instance,  the zero mode of $A_y$ has non-trivial $y$-dependence.
It is related to $\theta_H$ by
\beeq
g A_y =  \frac{k e^{2ky}}{e^{2\pi kR} - 1} \, \theta_H 
 \, \lambda^7 \equiv g A_y^c  ~~,~~
 \lambda^7 = 
 \begin{pmatrix} \\ &&-i \\ &i  \end{pmatrix}
 \label{RSzeromode}
\eneq

In the following discussions we concentrate on one particular component of
$A_y$ and $\theta_H$ as in (\ref{RSzeromode}), 
which corresponds to the neutral Higgs field in four
dimensions.  The argument can be generalized to theories with
multiple Higgs fields, which arise, for instance,   in theories on $M^4 \times T^n$.
In such cases one needs to consider a set of mutually independent 
$\theta_H$'s so that $F_{MN}=0$  even with nonvanishing $\theta_H$'s.

\bigskip
\leftline{\bf 2. Large gauge invariance}

A gauge transformation in the RS spacetime given by
\beeq
\Omega(y) = 
\exp \bigg( i n\pi \,  \frac{e^{2ky}-1}{e^{2\pi kR} -1} \, \lambda^7 \bigg) 
~~, ~~
(n: ~\hbox{an integer})
\label{largeGT1}
\eneq
preserves the boundary conditions (\ref{BC3}) but
shifts $\theta_H$ by $2\pi n$;
\beeq
\theta_H \go \theta_H' = \theta_H + 2 \pi n ~.
\label{largeGT2}
\eneq
The transformation is called a large gauge transformation.
It follows from the large gauge invariance that physical quantities 
such as the mass $m_H$ of the 4D Higgs boson are
periodic functions of $\theta_H$ with a period $2\pi$. 
In $M^4 \times (S^1 /Z_2)$ a large gauge transformation is
given by $\Omega = \exp \big( iny/R \cdot \lambda^7 \big)$, whereas
in $U(1)$ theory in $M^4 \times S^1$ it is given by 
$\Omega = \exp \big( iny/R  \big)$. @@

\bigskip
\leftline{\bf 3. Perturbation theory}
 
 In developing perturbation theory in the path integral formalism, we 
  separate $A_M$ into the classical part~$A^{\rm c}_M$ and the quantum 
part~$A^{\rm q}_M$;
\beeq
 A_M = A^{\rm c}_M+A^{\rm q}_M ~~. 
\eneq
In the $SU(3)$ model in the RS spacetime, $A^{\rm c}_y$ is defined
in (\ref{RSzeromode}) and $A_\mu^{\rm c}=0$.  
The quantum part $A^{\rm q}_M$ and fermion fields $\psi$ are
integrated in the path integral.   The background field gauge is specified with 
a gauge-fixing term $\Tr  f_{\rm g.f.} (A_M)^2$ where
\beqn
&&\hskip -1cm
 f_{\rm g.f.} = e^{2\sigma} \eta^{\mu\nu} 
 \D^{\rm c}_\mu A_\nu^{\rm q}
 + e^{2\sigma}  \D^{\rm c}_y
\big( e^{-2\sigma}A^{\rm q}_y \big) ~, \cr
\noalign{\kern 5pt}
 &&\hskip -1cm
\D^{\rm c}_M A^{\rm q}_N \equiv 
 \dd_M A^{\rm q}_N + ig \big[ A^{\rm c}_M , A^{\rm q}_N \big] ~. 
 \label{gf1}
\eeqn
$\sigma=0$ in the flat space.  The quadratic part of the  effective Lagrangian
which includes the gauge-fixing term and associated ghost term is simplified 
in the background field gauge. 

Under a large gauge transformation (\ref{largeGT1}),  $A_y^{\rm c}(\theta_H)$
is tranformed to 
$A_y^{\prime \,\rm c}(\theta_H) = A_y^{\rm c}(\theta_H + 2\pi)$, while
$A_M^{\rm q}$ to $A_M^{\prime \, \rm q} = \Omega A_M^{\rm q} \Omega^\dagger$.
It follows that $ f_{\rm g.f.} (A'_M) = \Omega  f_{\rm g.f.} (A_M) \Omega^\dagger$
so that the gauge-fixing term is invariant under the large gauge transformation. 
The invariance implies that the effective potential obtained in the new gauge 
$V_\eff (\theta_H + 2\pi)$ is the same as that in the old gauge  
$V_\eff (\theta_H )$;
\beeq
V_\eff (\theta_H + 2\pi) = V_\eff (\theta_H) ~.
\label{effV1}
\eneq
We stress that the periodicity in $\theta_H$ of $V_\eff$ is a 
consequence of the large gauge invariance.

The perturbation theory is developed with respect to $A_M^{\rm q}$ and $\psi$.
The total Lagrangian is decomposed as
\beqn
&&\hskip -1cm
\calL =\calL^{(2)} (A_M^{\rm q}, c, \bar c, \psi ; \theta_H) 
+ \calL^{(3)}_a (A_M^{\rm q}, c, \bar c, \psi ; g )  \cr
\noalign{\kern 5pt}
&&\hskip  -.2cm
+ \, \calL^{(3)}_b (A_M^{\rm q} ; g\theta_H ) 
+ \calL^{(4)} (A_M^{\rm q} ; g)
\label{perturbation1}
\eeqn
where $ \calL^{(2)}$ is bilinear in $A_M^{\rm q}$, $\psi$, and ghost fields 
$c, \bar c$.   $ \calL^{(2)}$ depends on $\theta_H$.  $ \calL^{(3)}_a$ and
$\calL^{(4)}$ are cubic and quadratic interactions, respectively,  
which are present  with vanishing $\theta_H$.  
 $ \calL^{(3)}_a = O(g)$ and $\calL^{(4)} = O(g^2)$.
The additional cubic interaction $\calL^{(3)}_b$ arises from the term 
$\onehalf g^2\Tr  [A_M, A_N] [A^M, A^N]$.    With (\ref{RSzeromode}) 
it becomes
\beeq
\calL^{(3)}_b = 2 g\theta_H \frac{k e^{2ky}}{e^{2\pi kR} -1}
\Tr [A_\mu^{\rm q} , A_y^{\rm q} ] [ A^{{\rm q} \, \mu} , \lambda^7 ] ~.
\label{perturbation2}
\eneq
It depends on $g$ and $\theta_H$ in the combination of $g\theta_H$.
In five dimensions the gauge coupling constant $g$ has a mass dimension
$-\onehalf$; $[g] = M^{-1/2}$.
 $ \calL^{(3)}_a$ and $\calL^{(4)}$ give non-renormalizable interactions, 
 while $\calL^{(3)}_b$ gives a super-renormalizable interaction.

\bigskip
\leftline{\bf 4. Finiteness in Abelian gauge theory}

With the property (\ref{effV1}),  $V_\eff (\theta_H)$ is expanded in
a Fourier series;
\beeq
V_\eff (\theta_H) = \sum_{n=-\infty}^\infty a_n \, e^{in \theta_H} ~.
\label{effV2}
\eneq
We are going to show  that 
$V_\eff (\theta_H) - a_0$ is finite to all order in
perturbation theory except at a discrete set of  values of $\theta_H$
in Abelian gauge theory,  where
$\calL^{(3)}_b$ and $\calL^{(4)}$ are absent.  The argument given in ref.\ 
\cite{YHscgt2} remains intact in this case.  
Expand $V_\eff (\theta_H)$ in $g$;
\beeq
V_\eff(\theta_H) = \sum_{\ell=0}^\infty g^{2\ell} ~ 
V_\eff^{(2\ell)} (\theta_H)  ~~.
\label{effV3}
\eneq
To each order in $g$ there are a finite number of bubble diagrams contributing
to $V_\eff^{(2\ell)} (\theta_H)$ in $U(1)$ theory.  
In each diagram $\theta_H$ appears in fermion propagators $S_F$.  In flat
space $S_F$ behaves as 
$[p_\mu \gamma^\mu + (n - a-\theta_H/2\pi)R^{-1} \gamma^5 - m_F]^{-1}$
where $a$ is a constant.  The periodicity in $\theta_H$ is recovered after summing 
over internal momentum indices $n$ in the fifth dimension.  
Here the translational invariance on $S^1$ is important.
One can expand  $V_\eff^{(2\ell)}$ in a Fourier series.
\beeq
V_\eff^{(2\ell)}  (\theta_H) = 
\sum_{n=-\infty}^\infty a_n^{(2\ell)} ~ e^{in\theta_H} ~~.
\label{effV4}
\eneq
We proceed to show that $V_\eff^{(2\ell)}  (\theta_H)  - a_0^{(2\ell)}$ is finite.  

Each diagram may be UV (ultraviolet)-divergent.
Now differentiate $V_\eff^{(2\ell)}$ with respect to $\theta_H$
sufficiently many times, say, $q$ times.  The divergence degree is lowered  
by $q$.  Since there are only a finite number of diagrams in $V_\eff^{(2\ell)}$,
$d^q V_\eff^{(2\ell)}/ d\theta_H^q$ becomes finite for sufficiently large $q$.
Hence $n^q a_n^{(2\ell)}$  becomes finite. 
It follows that $V_\eff^{(2\ell)} - a_0^{(2\ell)}$ is finite.

As shown in ref.\ \cite{YHscgt2}, the argument can break down at a discrete
set of values of $\theta_H$ when the fermion mass $m_F$ vanishes.  
Differentiation with respect to $\theta_H$ decreases the degree of UV divergence,
but increases the degree of IR (infrared) divergence when $m_F=0$ and
$a + (\theta_H/2\pi)=\hbox{an integer}$.   In even (odd) dimensions it induces
a singularity of the type of $\delta_{2\pi}(\theta_H)$ ($\ln \sin\theta_H$)
or its derivatives.   We note that the argument remains valid in arbitrary 
dimensions, in which case there are a multiple number of $\theta_H$'s.   
Thus we have proven a theorem.

\vskip 5pt
\noindent
{\bf [Theorem]~}  In $U(1)$ gauge theory defined in $M^4 \times T^n$ 
($n=1,2,3, \cdots$)  the $\theta_H$-dependent part of $V_\eff(\theta_H)$
is finite,  except at a discrete set of values of $\theta_H$,  in each order in 
perturbation theory.

\vskip 5pt

\leftline{\bf 5. Finiteness in non-Abelian gauge theory}

We proceed to the non-Abelian case, limiting ourselves to 
gauge theory in five dimensions.  In non-Abelian gauge theory propagators
of fermions, gauge fields, and ghost fields depend on $\theta_H$.  Further
$\calL^{(3)}_b$ and  $\calL^{(4)}$ are present,  the former of which  explicitly 
depends on $\theta_H$.  The presence of $\calL^{(3)}_b$ complicates 
the argument for the finiteness of the $\theta_H$-dependent part of 
$V_\eff (\theta_H)$.  Naively one might expand $V_\eff (\theta_H)$
in a power series of $g$ as in (\ref{effV3}).  If the expansion made sense,
the periodicity of $V_\eff^{(2\ell)}(\theta_H)$ would result and 
the argument presented in the Abelian case would apply.  However
this cannot be true as indicated by the following observation.  
Consider $O(g^2)$ corrections in $M^4 \times S^1$.  Among them
there is a two-loop diagram generated by two vertices of $\calL^{(3)}_b$,
which is proportional to $(g\theta_H)^2$.  Gauge field propagators also
depend on $\theta_H$.  After loop integrals and sums, it gives a contribution
of the form $(g\theta_H)^2 h(\theta_H)$ where $h(\theta_H)$ is 
periodic.  It follows that $V_\eff^{(2)}(\theta_H)$ would not be
periodic.  This implies that $V_\eff(\theta_H, g)$ is singular at $g=0$
in non-Abelian gauge theory.   The expansion (\ref{effV3}) would not be
valid. 

To distinguish contributions from $\calL^{(3)}_b$ and 
from $ \calL^{(3)}_a$ and $\calL^{(4)}$,
 we denote the coupling constant in the former by $\hat g$;
  $\calL^{(3)}_b (A_M^{\rm q} ; \hat g \, \theta_H )$.
 Let us  develop  perturbation theory in both parameters $g$ and $\hat g$, 
expanding  $V_\eff (\theta_H)$ as
\beeq
V_\eff(\theta_H) = \sum_{\ell=0}^\infty g^{2\ell} ~ 
\Big\{ V_a^{(2\ell)} (\theta_H) \, 
+  V_b^{(2\ell)} (\theta_H, \hat g \theta_H) \,  \big|_{\hat g = g}  \Big\} 
  ~~.
\label{effV5}
\eneq
$V_a^{(2\ell)} (\theta_H)$ contains a finite number of diagrams generated
by the vertices $\calL^{(3)}_a$ and  $\calL^{(4)}$, whereas 
$V_b^{(2\ell)} (\theta_H, \hat g \theta_H)$ is a sum of infinitely many diagrams
which are $O(g^{2\ell})$ and nonvanishing powers of $\hat g$ from $\calL^{(3)}_b$. 
$V_b^{(2\ell)}$ depends on $\theta_H$ through propagators and $\calL^{(3)}_b$. 
To have the periodicity in $\theta_H$, it seems necessary to have
contributions of all order in $\hat g$  in $V_b^{(2\ell)}$.  However, it is not
clear whether or not $V_a^{(2\ell)}  + V_b^{(2\ell)} $, for instance, is
periodic in $\theta_H$.

There exists special circumstance in five dimensions.  
$\calL^{(3)}_b$ gives super-renormalizable cubic
interactions.  The expansion parameter in $M^4 \times S^1$ or 
$M^4 \times (S^1/Z_2)$ is $\hat g \theta_H /R$, whereas 
$\hat g \theta_H  k$ in the RS warped spacetime.  The expansion parameter
has  mass dimension $+ \onehalf$.   This implies that the number of
UV divergent diagrams in $V_b^{(2\ell)}$ is finite.  
Each divergent diagram
contains a power  of $g \theta_H$ from vertices and a product of propagators.
The ultraviolet behavior of propagators is the same both in flat and RS spacetime.
Hence   $d_{}^q V_b^{(2\ell)} / d \theta_H^q$ becomes UV-finite 
by taking  sufficiently large $q$.

To prove the finiteness by generalizing the argument presented in the $U(1)$ case,
it is necessary to express $V_\eff(\theta_H)$ as a sum of 
gauge-invariant subsets of diagrams such that each subset 
is periodic in $\theta_H$.   It is likely that a whole set of diagrams in 
$V_b^{(2\ell)}$ is contained
in one of those subsets and that each subset contains only a finite
number of UV-divergent diagrams.  
The  argument breaks down in six or higher dimensions, 
where $\calL^{(3)}_b$ becomes marginal or non-renormalizable. 

Thus we conjecture the following.

\vskip 5pt
\noindent
{\bf [Conjecture]~}  In non-Abelian gauge theory defined in $M^4 \times S^1$,
$M^4 \times (S^1/Z_2)$, and the Randall-Sundrum warped spacetime, 
the $\theta_H$-dependent part of $V_\eff(\theta_H)$
is finite in each gauge invariant subset of diagrams,  except at a discrete set 
of values of $\theta_H$.

\bigskip
\leftline{\bf 6. Higgs boson mass}

The Higgs boson corresponds to four-dimensional  fluctuations of $\theta_H$.
Its mass $m_H$ is related to the curvature of the four-dimensional 
$V_\eff(\theta_H)$ at the minimum.\cite{HHHK, HNT2, Haba, HM}
For instance, in the $SU(3)$ model in the Randal-Sundrum warped spacetime
\beeq
m_H^2 = \frac{\pi g^2 R(e^{2\pi kR}-1)}{2k} \,
\frac{d^2 V_\eff}{d \theta_H^2} ~.
\label{HiggsMass1}
\eneq
In $U(1)$ gauge theory in $M^4 \times T^n$ a similar formula is obtained for the 
mass of 4D scalar fields arising from zero modes of extra-dimensional components 
of gauge potentials; $m_H^2 \sim g^2 R^2 (d^2 V_\eff /d\theta_H^2)$.
We note that $V_\eff (\theta_H) = M_{KK}^4 f(\theta_H)$ where $M_{KK}$ is the 
Kaluza-Klein mass scale and $f(\theta_H)$ is dimensionless.  

Although higher derivatives of $V_\eff(\theta_H)$ can be afflicted with infrared
divergence at a discrete set of values of $\theta_H$ as explained above,  the
global minimum in all non-Abelian models investigated so far is located at a regular point 
when the symmetry is dynamically broken.  In $U(1)$ theory with periodic (anti-periodic) 
fermions the global minimum occurs at $\theta_H=\pi$ (0), in either case of which 
$V_\eff(\theta_H)$ is regular at the minimum.  

Thus the finiteness of the $\theta_H$-dependent part of $V_\eff(\theta_H)$ implies the
finiteness of the Higgs boson mass.  One concludes that in $U(1)$ gauge theory defined
on $M^4 \times T^n$ ($n=1,2,3, \cdots$) the mass of the Higgs boson (4D scalar fields) 
is finite in each order in perturbation theory.  Radiative corrections are finite, being independent 
of the cutoff scale.  Recent two-loop analysis by Maru and Yamashita \cite{Maru1} 
supports the result in the present paper and  the earlier argument in ref.\ 
\cite{YHscgt2}.  The large gauge invariance plays a crucial role in the proof.

In non-Abelian gauge theory a proof for the finiteness of $V_\eff (\theta_H)$ is
incomplete. Based on the argument leading to the conjecture stated above, we
expect that the Higgs boson mass in non-Abelian gauge theory in
$M^4 \times S^1$, $M^4 \times (S^1/Z_2)$ and the Randall-Sundrum warped spacetime  
is finite in each order in the rearranged perturbation  theory where the large gauge 
invariance is maintained.
Although gauge theory in higher dimensions is non-renormalizable in perturbation
theory, the Higgs mass evaluated in the dynamical gauge-Higgs unification 
has well-defined  meaning free from the cutoff scale in the theory.      We  note that
it has been argued that non-Abelian gauge theory in five dimensions can be
defined in non-perturbative renormalization group approach \cite{Morris} and
in the  lattice formulation \cite{Irges}.    We shall come back to a more
detailed analysis of the non-Abelian case separately.


\vskip .5cm

\leftline{\bf Acknowledgments}
The author would like to thank Yutaka Sakamura and Toshio Nakatsu for
many enlightening discussions.
This work was supported in part by  Scientific Grants from the Ministry of 
Education and Science, Grant No.\ 17540257,
Grant No.\ 13135215 and Grant No.\ 18204024.

\vskip 2.cm

\def\jnl#1#2#3#4{{#1}{\bf #2} (#4) #3}

\def\Zphys{{\em Z.\ Phys.} }
\def\jssc{{\em J.\ Solid State Chem.\ }}
\def\jpsJ{{\em J.\ Phys.\ Soc.\ Japan }}
\def\ptps{{\em Prog.\ Theoret.\ Phys.\ Suppl.\ }}
\def\PTP{{\em Prog.\ Theoret.\ Phys.\  }}

\def\JMP{{\em J. Math.\ Phys.} }
\def\NPB{{\em Nucl.\ Phys.} B}
\def\NP{{\em Nucl.\ Phys.} }
\def\PLB{{\em Phys.\ Lett.} B}
\def\PL{{\em Phys.\ Lett.} }
\def\PRL{\em Phys.\ Rev.\ Lett. }
\def\PRB{{\em Phys.\ Rev.} B}
\def\PRD{{\em Phys.\ Rev.} D}
\def\PRe{{\em Phys.\ Rep.} }
\def\AP{{\em Ann.\ Phys.\ (N.Y.)} }
\def\RMP{{\em Rev.\ Mod.\ Phys.} }
\def\ZPC{{\em Z.\ Phys.} C}
\def\SCI{\em Science}
\def\CMP{\em Comm.\ Math.\ Phys. }
\def\MPLA{{\em Mod.\ Phys.\ Lett.} A}
\def\IJMPA{{\em Int.\ J.\ Mod.\ Phys.} A}
\def\IJMPB{{\em Int.\ J.\ Mod.\ Phys.} B}
\def\EPJC{{\em Eur.\ Phys.\ J.} C}
\def\PR{{\em Phys.\ Rev.} }
\def\JHEP{{\em JHEP} }
\def\cmp{{\em Com.\ Math.\ Phys.}}
\def\JPA{{\em J.\  Phys.} A}
\def\JPG{{\em J.\  Phys.} G}
\def\NJP{{\em New.\ J.\  Phys.} }
\def\CQG{\em Class.\ Quant.\ Grav. }
\def\ATMP{{\em Adv.\ Theoret.\ Math.\ Phys.} }
\def\ibid{{\em ibid.} }

\renewenvironment{thebibliography}[1]
         {\begin{list}{[$\,$\arabic{enumi}$\,$]}  
         {\usecounter{enumi}\setlength{\parsep}{0pt}
          \setlength{\itemsep}{0pt}  \renewcommand{\baselinestretch}{1.2}
          \settowidth
         {\labelwidth}{#1 ~ ~}\sloppy}}{\end{list}}

\def\reftitle#1{}                

\end{document}